\documentclass[prl,aps,twocolumn,groupedaddress,floats,showpacs,final]{revtex4}
\usepackage{graphicx}
\usepackage{dcolumn}
\usepackage{bm}
\usepackage{color}
\usepackage[caption=false]{subfig}

\definecolor{DarkGreen}{rgb}{0,0.7,0.08} 
\definecolor{Grey}{rgb}{0.5,0.5,0.5}
\definecolor{Red}{rgb}{0.8,0.3,0.3}
\definecolor{Green}{rgb}{0.1,0.8,0.1}
\definecolor{Blue}{rgb}{0.1,0.1,0.8}
\newcommand{\oldtext}[1]{}

\begin{document}

\title{
Coexistence, interfacial energy and the fate of  microemulsions  in  2D dipolar bosons}
\author{Saverio Moroni$^1$ and Massimo Boninsegni$^{2}$}
\affiliation{$^1$ {SISSA Scuola Internazionale Superiore di Studi Avanzati and DEMOCRITOS National 
Simulation Center,
Istituto Officina dei Materiali del CNR Via Bonomea 265, I-34136, Trieste, Italy}}
\affiliation{$^2$ Department of Physics, University of Alberta, Edmonton, Alberta, Canada, T6G 2E1}
\date{\today}                                           

\begin{abstract}
The superfluid-crystal quantum phase transition of a  system of purely repulsive dipolar bosons in two dimensions is studied by 
Quantum Monte Carlo simulations at zero temperature. We determine  freezing and melting densities, and estimate the energy per unit length of a macroscopic interface separating the two phases. The results rule out the microemulsion scenario for any physical realization of this system, given the exceedingly large predicted size of the bubbles.

\end{abstract}

\pacs{05.30.Jp, 05.30.Rt, 68.05.Gh}
\maketitle

The phase diagram of an assembly of spin-zero Bose particles in two dimensions (2D), interacting via the purely repulsive pair-wise potential $V(r)= D/r^3$ has been the subject of much theoretical investigation over the past decade \cite{general,general2}.
The quantum-mechanical Hamiltonian, in dimensionless form, reads as follows:
\begin{equation}
\hat H=-\frac{1}{2}\sum_{i=1}^N
\nabla^2_i + \sum_{i <j}\frac{1}{|{\bf r}_i-{\bf r}_j|^3}
\label{ham}
\end{equation}
All lengths are expressed in terms of $a\equiv mD/\hbar^2$, $m$ being the particle mass \cite{note0}, whereas the energy unit is $\epsilon_0\equiv
\hbar^2/ma^2 = D/a^3$.  The physics of this model at temperature $T=0$ is controlled by a single parameter, i.e.,  the density $\rho$. \\ \indent
The many-body Hamiltonian (\ref{ham})  is of fundamental interest for a number of reasons, from the effect of the interactions  (neither  short nor truly long ranged in 2D) on the superfluid properties \cite{filinov}, to the microscopic character of the quantum (i.e., $T=0$) phase transition from a superfluid to an insulating crystal, which remains to be elucidated. \\ \indent
A general argument has been proposed \cite{spivak} to the effect that no conventional first order phase transition can occur in such a system, in 2D. Specifically, in the presence of an interaction falling off as $1/r^3$, the coexistence of two phases of different density (crystal and superfluid)  separated by a macroscopic interface is  energetically unfavorable. The system can lower its energy by forming a “microemulsion”, featuring large solid clusters (or ``bubbles")  floating in the superfluid. At low temperature, bubbles are predicted to arrange themselves into a lattice superstructure, owing to their large mass, in essence giving rise to a ``supersolid" phase \cite{rmp}.
\\ \indent
This intriguing prediction could in principle be tested experimentally, as there exist a number of possible physical realizations of (\ref{ham}). For example, ultracold assemblies of Rydberg-excited atoms \cite{gallagher} can be confined to 2D  by means of an external harmonic trap; upon aligning their electric dipole moments in the direction perpendicular to the plane of confinement by means of an external electric field, the interaction between any two particles is $V(r)=D/r^3$, where $D$ in this case is proportional to the square of the dipole moment.  However, (\ref{ham}) is also apt to describe a system of indirect excitons in 
semiconductor quantum wells \cite{qw}.
\\ \indent
The ground state phase diagram of (\ref{ham}) has been studied by computer simulations \cite{pupillo,bor,mora}, which have identified a 
low-density superfluid and a high-density crystalline phase, but yielded no evidence  of the microemulsion proposed in Ref. \cite{spivak}.
Indeed, there is an aspect of the argument furnished therein that has not yet been fully clarified (or even addressed) quantitatively, despite its obvious experimental (or simulational) relevance, namely that of  the typical size $R$ of the bubbles in the microemulsion. It can be shown that \cite{vanderbilt}
\begin{equation}\label{ng}
R = d\ {\rm exp} \biggl (\frac{\gamma_b}{\gamma_d}\biggr)
\end{equation}
where $d$ is a characteristic length, $\gamma_d = \epsilon_0 a^3 (\rho_S-\rho_L)^2$, $\rho_S$ and $\rho_L$ being the melting and freezing densities, and $\gamma_b$ is the energy per unit length of a macroscopic interface. The length $d$ depends on the geometry and on the specific physical settings, but can be generally expected to be  a few times the average interparticle spacing \cite{vanderbilt}.
\\ \indent
To our knowledge, no estimates are presently available of either $\gamma_b$ or $\gamma_d$ (in any case, not of accuracy sufficient to estimate $R$ reliably); because of its exponential dependence on the ratio of these two quantities, $R$ could be conceivably extremely large, rendering the microemulsion scenario  of academic interest only. Indeed, this was suggested to be the case for a 2D Coulomb system \cite {casula}, for which 
{\color{blue}
} a similar prediction had been made
 \cite{spivak2,ortix}.
\\ \indent
In this Letter, we provide robust numerical evidence that even for the ``marginal" $1/r^3$ interaction,  the size of the system needed to observe the microemulsion greatly exceeds anything even imaginable, much less experimentally accessible, for any realistic value of $a$. Specifically, by means of ground state Quantum Monte Carlo simulation we compute $\rho_S$ and $\rho_L$, and estimate $\gamma_b$. Even making allowance for the statistical and systematic uncertainties affecting our calculation, the results show that $\gamma_b$ is as much as {\em four orders of magnitude} greater than $\gamma_d$, largely due to the remarkable narrowness of the coexistence region. Thus, for all practical purposes the quantum phase transition occurring in this system can be regarded as a conventional first order one.
\\ \indent
We reached the above conclusion by studying the zero-temperature phase diagram of (\ref{ham}) by means of computer simulations, based on the Path Integral Ground State (PIGS) method \cite{sarsa}, which is particularly well suited to investigate the ground state of Bose systems. It is essentially a variational approach \cite{bruttoceffo}, in which an arbitrarily accurate approximation to the ground state wave function is obtained as
\begin{equation}\label{prime}
\Psi(\Lambda) =  {\rm exp}(-\Lambda\hat H) \Psi_T
\end{equation}
as $\Lambda\to\infty$,  $\Psi_T$ being a trial wave function. In this work, we use
\begin{equation}\label{wf}
\Psi_T={\rm exp}\biggl [ -\sum_{i<j} u(r_{ij})\biggr ]\ \times\ {\rm exp}\biggl [-\alpha\sum_{i=1}^N (|{\bf r}_i-{\bf b}_i|^2)\biggr ] 
\end{equation}
Here, the Jastrow pseudopotential $u$ is optimized as described in Ref. \cite{fantoni}, whereas  ${\bf b}_1,... {\bf b}_N$ are the sites of a 2D triangular lattice at which particles are ``pinned", if the variational parameter $\alpha\ne 0$, in which case the wave function explicitly breaks translational invariance, i.e., corresponds to a crystalline ground state. On the other hand, if $\alpha=0$, {\em ansatz} (\ref {wf}) is translationally invariant, and apt to describe  a superfluid. 
\\ \indent
In principle, in the limit $\Lambda\to\infty$ the PIGS algorithm should extract the true ground state wave function for a Bose system regardless of which (positive-definite) initial trial wave function $\Psi_T$ is chosen.
In practice a finite projection time $\Lambda$ is used, hence the  physics of the projected state generally reflects that of the trial wave function; thus, the energy expectation value is lower at low density  on setting $\alpha=0$  in $\Psi_T$, as the system is in the superfluid phase, whereas the crystalline {\em ansatz} ($\alpha\ne 0$)  yields a lower energy at high density. 
\\ \indent
On approaching the coexistence region from the high and low density sides,  the pressure $P$ and  chemical potential $\mu$ for the two phases become equal at the two densities $\rho_S$ (melting) and $\rho_L$ (freezing), i.e., the condition of phase equilibrium.
In the case of power law type interactions,  pressure and chemical potential can be  obtained from the total and potential energy per particle, as one can show using the virial theorem \cite{imada}. In particular, one has
\begin{equation}
\frac{P}{\rho} = e + \frac{1}{2}v
\end{equation}
and $\mu=e+P/\rho$, where $e$ and $v$ are the total and potential energy per particle, which are directly accessible by simulation.
\\ \indent
We carried out simulations of  systems
comprising up to $N$=400 particles, enclosed in a rectangular  cell capable of accommodating a perfect triangular lattice \cite{note}, with periodic boundary conditions. A typical value of the projection time utilized is $\Lambda \sim 1/(\rho a^2\epsilon_0)$; we used the primitive approximation for the short-time propagator, with a  time step $\tau \sim 10^{-3}\Lambda$. 
\\ \indent
An important aspect of the calculation, given that the interaction among particles is not short-ranged, consists of estimating the  contribution $\Delta v$ to the potential energy per particle arising from particles outside the largest distance $r_c$ allowed by the simulation cell. We do that by fitting the tail of the 
pair correlation function $g(r)$ for the largest system size to 
{\color{blue}
}
a damped oscillation around unity \cite{bruttoceffo2},
and obtain $\Delta v$ as 
$\pi\rho \int_{r_c}^\infty dr \ r\ g(r)\ r^{-3}$.
\\ \indent
We assess combined statistical and systematic errors of our energy estimates, due to a finite projection time, finite time step and finite system size, to amount to no more than  a fraction $5\times 10^{-5}$ of the energy value. 
The only previous study with which we can directly compare our results is that of  Astrakharchik {\it et al.}, who studied the ground state of (\ref {ham}) using the  Diffusion Monte Carlo (DMC) method \cite{bor}. Our energy values extrapolated to the thermodynamic limit, as a function of $\rho$, are consistently, significantly {\it lower} than theirs, most notably in the crystalline phase \cite{noteschifo}. \\
\begin{figure}[h]
\centerline{\includegraphics[height=2.3in]{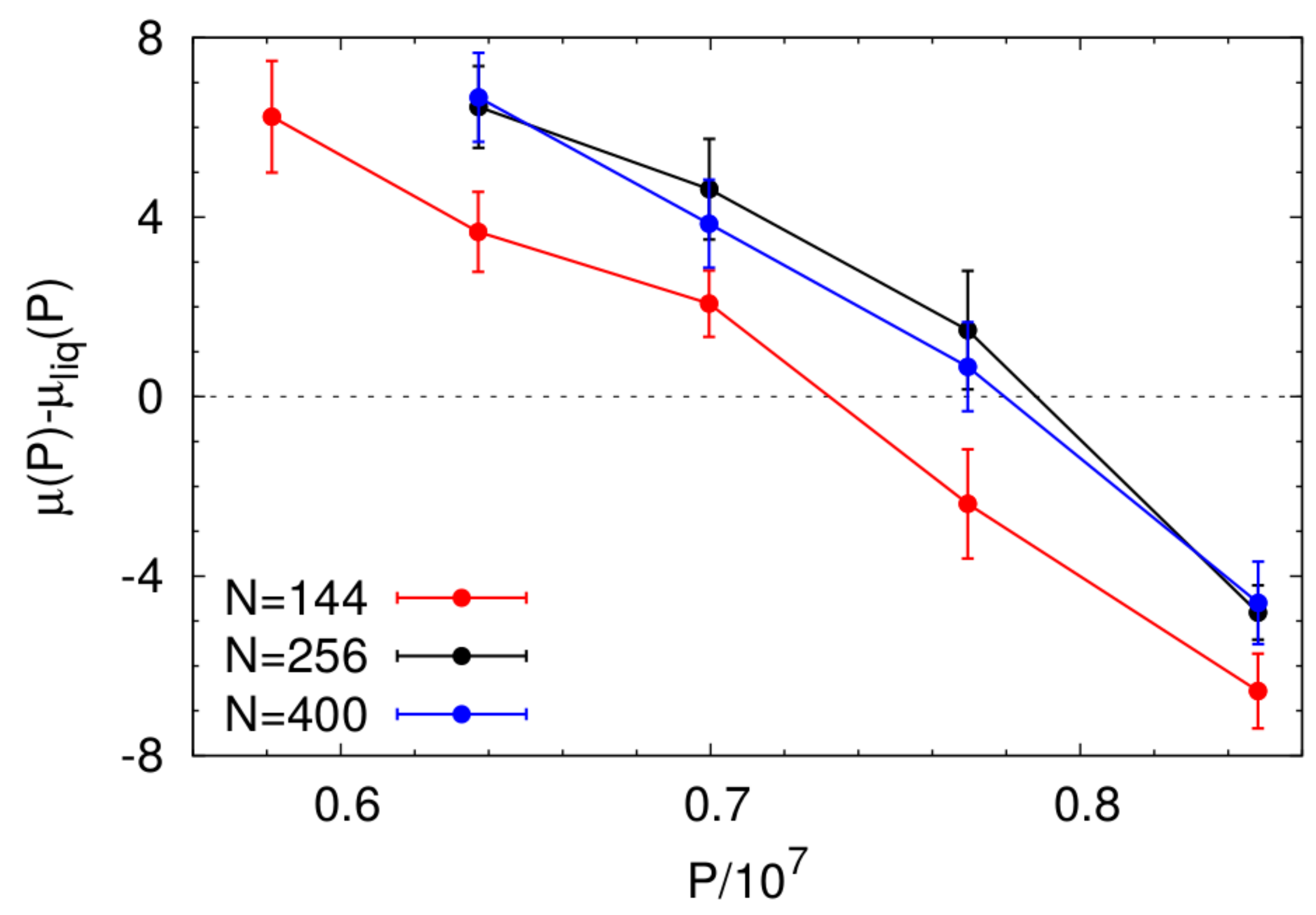}}
\caption{(Color online). Ground state chemical potential $\mu$ as a function of pressure computed in the solid phase for systems comprising $N=144$ (red), 256 (black) and 400 (blue) particles. Dotted line is a fit to the values of $\mu$ for the liquid for the largest system size, taken as reference
\cite{note2}.
Statistical errors on the values of the pressure are smaller than symbol size. 
Straight lines connecting points are only a guide to the eye. }\label{muofp}
\end{figure}
\\ \indent 
Fig. \ref{muofp} shows computed values of the chemical potential and pressure for the two phases, for three different system sizes. The  values for the liquid phase for the largest system size considered here  are taken as reference for convenience
\cite{note2}. 
Although 
the size dependence of the results is still noticeable at $N=144$,
the estimated intersection of the chemical potential of the two phases occurs at the same pressure,
within the statistical uncertainties of the calculation, for a system with $N=256$ and $N=400$. 
Using the results for the largest system, and taking into account the statistical 
error on the chemical potential in the liquid phase, we locate the transition 
at $(7.8\pm0.3)\times 10^6\ \epsilon_0 a^{-2}$.
\begin{figure}[h]
\centerline{\includegraphics[height=2.2in]{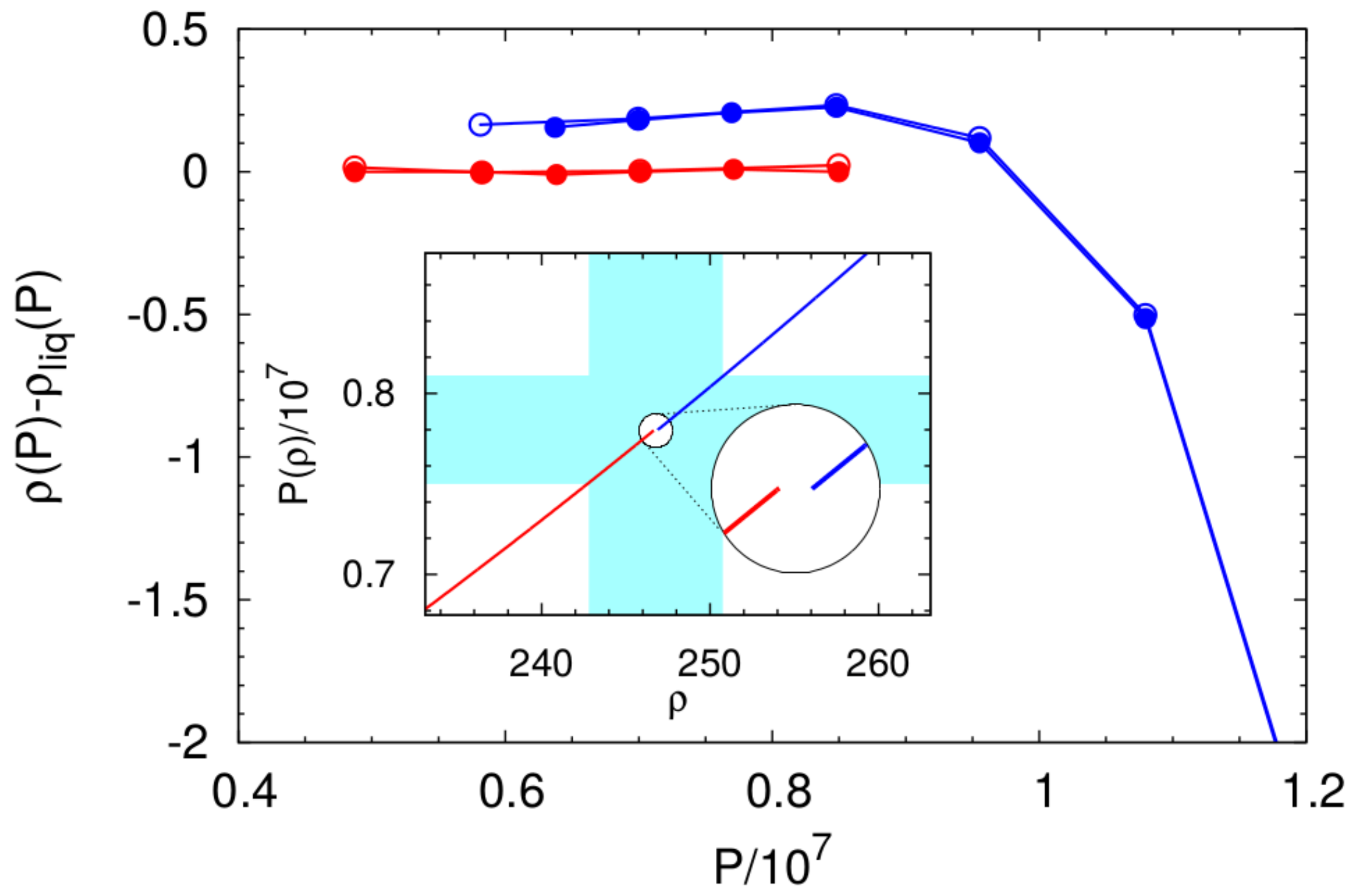}}
\caption{(Color online). Density versus pressure for both the crystal (blue symbols and lines) and the liquid
(red symbols and lines) phases. Open symbols refer to results for a system with $N=144$ particles, filled symbols to one with $n=400$ particles. 
Statistical errors are smaller than symbol size.
Values in the liquid phase for the largest system are taken as reference. {\it Inset}: expanded view of the region in which phase transition is estimated to occur (shaded area). Blowup shows density jump corresponding to liquid-solid transition.}\label{rhoofP}
\end{figure}
\\ \indent
Having estimated the coexistence pressure, melting and freezing densities can be deduced from the values of pressure versus density for both phases, shown in Fig. \ref{rhoofP}. The first observation is that the coexistence region is very narrow. This conclusion is rather robust, as the data in the figure show that the difference between liquid and solid density is 
almost constant in a fairly wide region around the location of the
phase transition.
Specifically, while  $\rho_S, \rho_L = 
(247\pm 4)\
a^{-2}$ (shaded region in inset), we can state with high confidence that $0.17\ a^{-2} \le (\rho_S-\rho_L) \le 0.24\ 
 a^{-2}$, i.e., $0.03\ \epsilon_0 a^{-1}\le \gamma_d \le 0.06\
\epsilon_0 a^{-1}$ (see eq. \ref{ng}). Our determined freezing and melting densities are not inconsistent with results of previous studies \cite{pupillo, bor,mora}, although  the very large uncertainties quoted therein render a direct comparison scarcely meaningful. 
\begin{figure}[h]
\centerline{\includegraphics[height=2.0in]{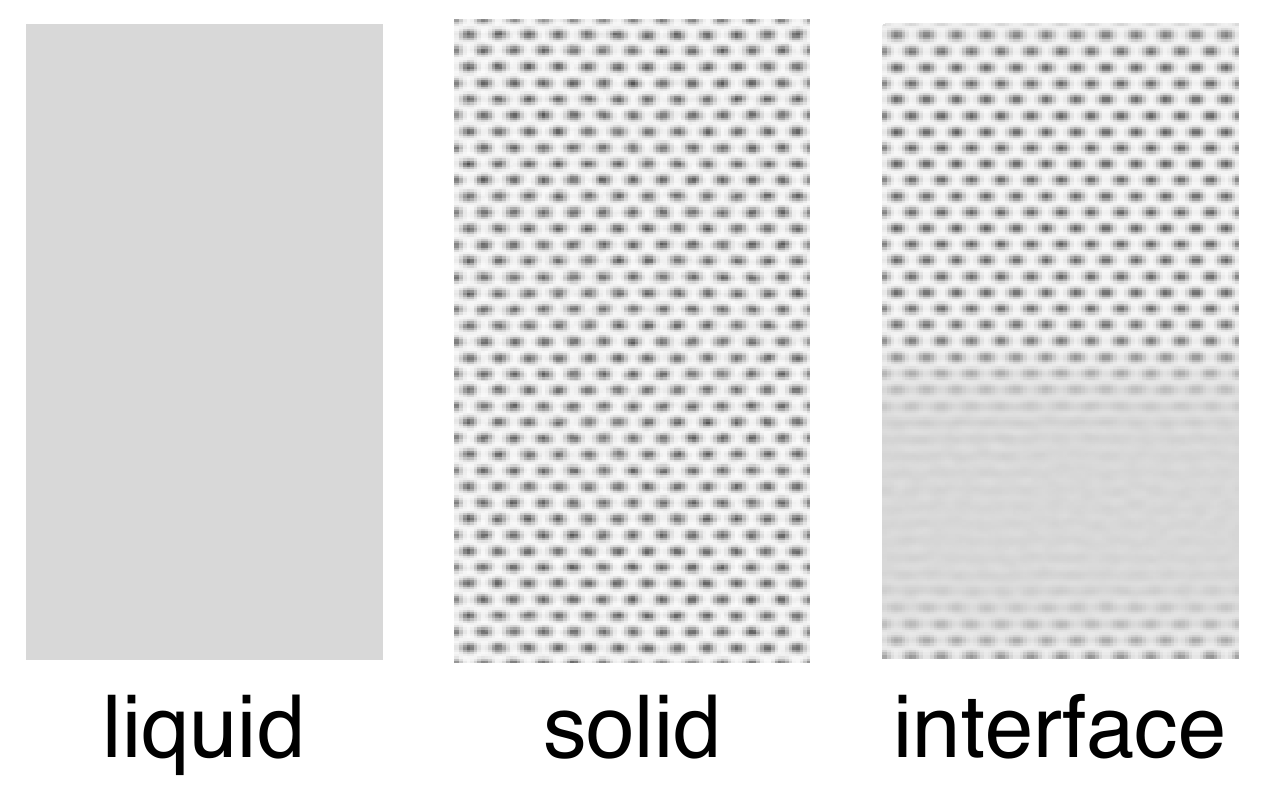}}
\caption{One-body density profiles pertaining to simulations of the system in the liquid (left) and solid (middle) phases, as well as of an interface between the two phases (right).}\label{inter}
\end{figure}
\\ \indent
The computed values of $\rho_L$ and $\rho_S$ suggest $d \sim a$ in Eq. \ref{ng};
 however,  assessing the characteristic size $R$ of a ``bubble" in the speculated microemulsion requires knowledge of the energy per  unit length $\gamma_b$ of a macroscopic interface separating the two phases at coexistence.  In order to obtain an estimate for $\gamma_b$, we  follow a procedure similar to that of Ref. \cite{pederiva}, i.e.,  carry out a separate simulation of an actual interface. \\ \indent 
The setup is shown in fig. \ref{inter}. We use an elongated cell of sides $L$ and $L^\prime \approx 2.85\ L$, and divide it into two regions (rightmost panel), one occupied by liquid, the other by solid, separated by an interface of length $L$ \cite{notez}.
The density $\rho_0$ is chosen for simplicity to be the same for both phases. We simulated systems with a total number of particles $N$ between 168 and 474,  for values of $\rho_0$ inside and near the coexistence region.
We stabilize the interface by making the parameter $\alpha$ in the wave function (\ref {wf}) dependent on position, i.e., we set it to zero in half of the cell, where particle are allowed to wander about, and to a finite value in the other half, where particles are pinned at lattice sites. 
\\ \indent
Concurrently, we also performed simulations with the same geometry but with only one of the two phases, and the same density $\rho_0$ (leftmost and middle panel of fig. \ref{inter}). We project in all cases for an imaginary time interval $\Lambda=5\times 10^{-3}\ \epsilon_0^{-1}$,
 and estimate the energy of the interface as
\begin{equation}\label{differenza}
L\gamma_b = \frac{1}{2}\biggl [ E_I - N_S e_S-(N-N_S)e_L\biggr]
\end{equation}
where the factor 1/2 comes from the presence of two interfaces \cite{notez}, $E_I$ is the total energy of the simulated system with the interface, and $e_L$ ($e_S$) is the energy per particle in the liquid (solid) phase, computed in the separate simulations of homogeneous systems. For this part of the study, given the inhomogeneity of the system and the shape of the simulation cell, we  computed the contribution to the potential energy associated to particles beyond the maximum distance  allowed by the cell  also by means of an explicit summation of  contributions from particle images in adjacent cells, up to a maximum distance of 10 $L$;  contribution from particles at greater distances was estimated by assuming $g(r)=1$ and integrating from 10 $L$ to infinity.   As it turns out, this procedure yields results compatible with those furnished by that described above, utilized for the homogeneous phases.
\\ \indent
The value of $\gamma_b$ obtained in this way is 1600 $\epsilon_0 a^{-1}$, at $\rho_0=0.247\ a^{-2}$; our assessment of combined statistical and systematic errors is at no more than 10\% of this value. 
This results into an order of magnitude estimate of the ratio $\gamma_b/\gamma_d \sim 10^4$, which in turn makes $R$ infinite for all practical purposes, based on Eq. \ref{ng}.\\ \indent
A few comments are in order:
\\
1) Only a reliable order-of-magnitude estimate for $\gamma_b$ is needed, given the narrowness of the coexistence region. It is interesting to note that if the coexistence region were as wide as allowed, for instance, by the calculations of Refs. \cite{pupillo, bor} ($\sim 30-100\ a^{-2}$), then our estimate for $\gamma_b$ would lead to an entirely different physical conclusion, as $R\sim d$ in that case.
\\
2) The methodology adopted here is in principle unbiased, even though it does require an input trial wave function.  
The most important sources of systematic error of this calculation are the
finite size of the simulated system and the finite projection time. Simulation
of a  system with $N=168$, with eight rows of 6 particles each pinned at lattice sites, yields the same interface energy per unit length obtained for a system of $N=474$ particles, within statistical uncertainties. The fact that the estimate of $\gamma_d$ does not change on (almost) tripling the
system size, constitutes in our view strong evidence of its robustness. Furthermore, full extrapolation in projection time for the smaller system shows that the bias incurred at $\Lambda=5\times 10^{-3} \epsilon_0^{-1}$ is less than 
$40\ \epsilon_0a^{-1}$.
\\ 
3) The energy of the interface $E_I$ is approximately 0.03\% of the total energy of the system. Its value is insensitive to the relative numbers of particles in the two phases, because the energies $e_S$ and $e_L$ are very close at the density $\rho_0$ considered here, as well as at coexistence. It is interesting to compare it with that obtained in Ref. \cite{pederiva} for 3D $^4$He, where it was found that the energy per atom of the interface (assumed to consist of a single atomic plane) is worth approximately 15\% of the atomic kinetic energy in the superfluid, at coexistence. The same estimate yields 5\% in our case.
\begin{figure}[h]
\centerline{\includegraphics[height=2.4in]{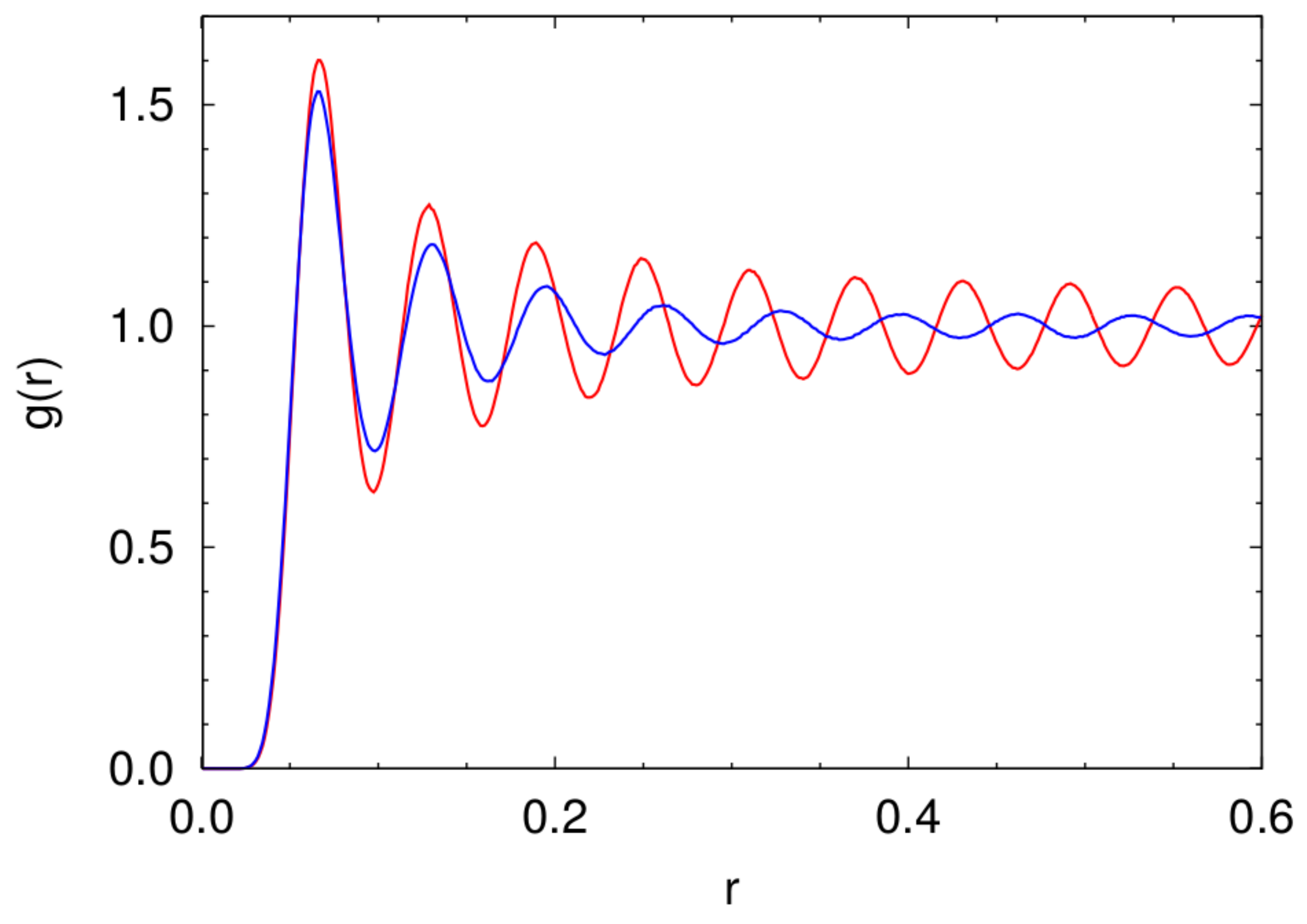}}
\caption{(Color online). Pair correlation functions $g(r)$ for the liquid (blue) and solid(red) phases, computed
at $\rho_0=237\ a^{-2}$, on a system of $N$=400 particles. Statistical errors are not visible on the scale of the figure. }\label{gr}
\end{figure}
\\
4) The assumption of equal density for the coexisting phases is made for convenience, and is justified by the narrowness of the coexistence region. It allows one to isolate the energetic contribution of the surface tension, i.e., to  estimate $\gamma_b$ through Eq.  \ref{differenza}. The estimate of $\gamma_b$ arrived at through the procedure outlined above is rather insensitive to the value of $\rho_0$ chosen; for example, on performing the calculation with $\rho_0=237\ a^{-2}$ we obtain $\gamma_b\approx 1500\ \epsilon_0 a^{-1}$.
The pair correlation function $g(r)$ for the two coexisting phases is shown in Fig. \ref{gr}, for the case $\rho_0=237 \ a^{-2}$. 
\\ \indent
Summarizing, by means of accurate Quantum Monte Carlo simulations we have determined melting and freezing densities of a two-dimensional system of spin-zero bosons interacting via a purely repulsive potential of the form $V(r)=D/r^3$, and evaluated the energy per unit length of a macroscopic interface separating coexisting superfluid and crystalline phases. The width of the coexistence region is remarkably small, of the order of 0.01\% of the freezing (melting) density. The  most important implication is that the characteristic size of the bubbles that should constitute the microemulsion, a physical scenario proposed in Ref. \cite{spivak} as energetically competitive with simple coexistence,  exceeds anything experimentally accessible, given the assessed value of the interfacial energy. \\ \indent
Thus, for all practical purposes a conventional first-order phase transition between the superfluid and crystalline phases is all that can be observed either experimentally or in simulations, for a system of this kind.
More generally,  any evidence of ``bubble" phases in numerical simulations should be carefully assessed, especially when not supplemented by quantitative estimates of melting and freezing density, as well as of the interfacial energy.
\\ \indent
This work was supported in part by the Natural Science and Engineering Research Council of Canada. MB gratefully acknowledges the hospitality of the International School for Advanced Studies in Trieste, as well as of the ETH Z\"urich, where part of this work was carried out.


\begin{thebibliography}{99}
\bibitem{note0}
The parameter $a$ can span a rather wide range, from $\sim$ 1 \AA\ for electronic systems, to  $\sim$ 1 m for Rydberg atoms, 
as one can ascertain by expressing it as $\kappa^2 m^\star a_0$, where $\kappa$ is the electric dipole moment in Debye, $m^\star$ is the mass of the particles in units of the electron mass, and $a_0$ is the Bohr radius.
\bibitem{note}
The shape of the cell does not obviously affect simulations of the system in the superfluid phase.
\bibitem{note2}
In the liquid phase the size dependence of the energy is slightly weaker than
in the solid. Correspondingly, the values of $\mu$ are within error bar 
for $N=144,$ 256 and 400.
\end{thebibliography}
\end{document}